\documentclass[aps,prc,preprint,groupedaddress,showpacs]{revtex4}
\usepackage[dvipdfmx]{graphicx}
\usepackage{amsmath}

\begin{document}
\title{Changes in rotational characters of one- and two-phonon $\gamma$-vibrational bands in $^{105}$Mo}

\author{Masayuki Matsuzaki}
\email[]{matsuza@fukuoka-edu.ac.jp}
\affiliation{Department of Physics, Fukuoka University of Education, 
             Munakata, Fukuoka 811-4192, Japan}

\date{\today}

\begin{abstract}
The $\gamma$ vibration is the most typical low-lying collective motion 
prevailing the nuclear chart. But only few one-phonon rotational bands in odd-$A$ nuclei 
have been known. Furthermore, two-phonon states, even the band head, have been observed 
in a very limited number of nuclides not only of odd-$A$ but even-even. 
Among them, that in $^{105}$Mo is unique in that Coriolis effects are 
expected to be stronger than in $^{103}$Nb and $^{105}$Nb on which theoretical studies 
were reported. Then the purpose of the present work is to study $^{105}$Mo paying 
attention to rotational character change of the one-phonon and two-phonon bands. 
The particle-vibration coupling model based on the cranking model and 
the random-phase approximation is used to calculate the vibrational states in rotating 
odd-$A$ nuclei. 
The present model reproduces the observed yrast zero-phonon and one-phonon 
bands well. Emerging general features of the rotational character change from low spin to high 
spin are elucidated. In particular, the reason why the one-phonon band does not exhibit 
signature splitting is clarified. The calculated collectivity of the two-phonon states, 
however, is located higher than observed. 
\end{abstract}

\maketitle

\section{Introduction}

Based on the long history of physics of the atomic nucleus as a finite quantum many-body 
system bound by the strong interaction, nowadays, on the one hand interactions between 
two, or among three or more, nucleons can be related to quantum chromodynamics and 
the interaction thus obtained can be used as an input to large-scale shell-model 
calculations. On the other hand, from the mean-field picture, 
density-functional theories that describe ground states of nuclides of wide range 
in the nuclear chart are developing. For a class of excited states, the 
generator-coordinate method is used. Applications of such a framework to light 
odd-$A$ nuclei are just begun~\cite{BABH}. 

Aside from these progresses, traditional effective models are still indispensable to 
describe detailed properties of collective excitations. 
In finite many-body systems, individual particle motions and collective motions 
are of similar energy scales hence couple to each other. In addition, they are sensitive 
to the numbers of constituents or the shell filling. Collective vibrations are 
not necessarily harmonic oscillations and whether multiple excitations exist or not 
is not trivial. 

One of the typical collective motions in medium-heavy nuclei is the $\gamma$ vibration. 
But its double excitation has been known in a limited number of nuclides such as 
$^{168}$Er~\cite{Er168}, $^{166}$Er~\cite{Er166_1,Er166_2}, $^{164}$Dy~\cite{Dy164}, 
$^{232}$Th~\cite{Th232_1}, $^{106}$Mo~\cite{Mo106}, $^{104}$Mo~\cite{Mo104}, and 
$^{138}$Nd~\cite{Nd138} in even-even nuclei. Among them, $^{168}$Er, firstly observed one, 
was studied in terms of various theoretical approaches as reviewed in Ref.~\cite{MM1}. 

In odd-$A$ nuclei, the first observation of the two-phonon $\gamma$ vibration was made in 
$^{105}$Mo~\cite{Mo105}, on which any theoretical studies has not been reported 
to the author's knowledge, and to which the present study is devoted. Then similar excitations 
were observed in $^{103}$Nb~\cite{Nb103} and $^{107}$Tc~\cite{Tc107}. 
The first realistic calculation on $^{103}$Nb~\cite{SBSP} was performed 
in terms of the triaxial projected shell model. After that the present author made a 
calculation on this nuclide in terms of the particle-vibration coupling model in Ref.~\cite{MM1}. 

Quite recently, two-phonon states very similar to those in $^{103}$Nb were observed in 
$^{105}$Nb~\cite{Nb105}. In these isotopes, candidates of three-phonon states that fed 
two-phonon bands strongly were also indicated. 
If their property is confirmed, they are the first three-phonon 
excitation in deformed nuclei, and may indicate that there is a mechanism that makes odd-$A$ 
nuclei more favorable for realizing three-phonon $\gamma$ vibrations than even-even nuclei, 
although spectra of odd-$A$ nuclei are thought to be more complex than those of even-even 
nuclei in general. The present author studied in Ref.~\cite{MM2} these three-phonon candidates 
with invoking a method to calculate interband $B(E2)$s based on the generalized intensity 
relation~\cite{SN}. The result is promising but still not definitive. 

Returning to two-phonon states, the odd particle that couples to them in $^{103}$Nb and $^{105}$Nb 
is in the $\pi[422]\,5/2^+$ orbital originating from the $g_{9/2}$ subshell on which Coriolis effects 
are not very strong. Then states at finite rotation still can be approximately classified 
in terms of $K$, the projection of the angular momentum to the $z$ axis. 
In contrast, the odd particle in $^{105}$Mo is in the $\nu[532]\,5/2^-$ orbital originating from the 
$h_{11/2}$ subshell on which Coriolis effects are stronger. Accordingly character of rotational 
band members would be different. This point is the main concern of the present work. 

Throughout this paper the $\hbar=1$ unit is used. 

\section{The model and parameters}

The particle-vibration coupling (PVC) model based on the cranking model and the random-phase 
approximation (RPA) is used for calculating eigenstates of odd-$A$ nuclei. 
In the first step, the one-dimensionally cranked Nilsson plus BCS one-body Hamiltonian 
is diagonalized, 
\begin{equation}
\begin{split}
h'&=h-\omega_\mathrm{rot}J_x \\
  &=\sum_{\mu}e'_{\mu}a^\dagger_{\mu} a_{\mu}+\sum_{\bar\mu}e'_{\bar\mu}a^\dagger_{\bar\mu} a_{\bar\mu}
   +\mathrm{const} ,
\end{split}
\end{equation}
where $h$ is the standard Nilsson plus BCS Hamiltonian. 
This gives a set of quasiparticle (qp) states. Second, I apply the 
RPA to the residual two-body pairing plus doubly-stretched quadrupole-quadrupole 
interaction between qps, 
\begin{equation}
\begin{split}
H_\mathrm{int}&=
-\sum_{\tau=1,2} G_\tau \tilde P_\tau^\dagger \tilde P_\tau
-\frac{1}{2}\sum_{K=0,1,2} \kappa_K^{(+)} Q_K''^{(+)\dagger} Q_K''^{(+)} 
-\frac{1}{2}\sum_{K=1,2} \kappa_K^{(-)} Q_K''^{(-)\dagger} Q_K''^{(-)} \\
              &=\sum_{n}\omega_n X_n^\dagger X_n .
\end{split}
\end{equation}
This gives many RPA modes. Cranking and RPA calculations are done 
in the five major shells, $N_\mathrm{osc}=2$ -- 6 for the neutron and 1 -- 5 for the proton. 
Among RPA modes, I choose the $\gamma$-vibrational phonons, $n=\gamma$ and $\bar\gamma$, 
with signature $r=\exp{(-i\pi\alpha)}=1$ and $-1$, respectively, 
which have outstandingly large $K=2$ transition amplitudes. 
Regarding them as elementary excitations, I diagonalize the PVC Hamiltonian,
\begin{equation}
\begin{split}
H_\mathrm{couple}(\gamma)
     &={\sum_{\mu\nu}}\Lambda_{\gamma}(\mu\nu)
       \left(X_{\gamma}^\dagger a_\mu^\dagger a_\nu
           + X_{\gamma}         a_\nu^\dagger a_\mu\right) \\
     &+\sum_{\mu\bar\nu}\Lambda_{\bar\gamma}(\mu\bar\nu)
       \left(X_{\bar\gamma}^\dagger a_\mu^\dagger a_{\bar\nu}
           + X_{\bar\gamma}         a_{\bar\nu}^\dagger a_\mu\right) \\
     &+\mbox{sig. conj.} 
\end{split}
\end{equation}
These calculations are performed in terms of signature-classified bases, and all the 
resulting quantities are given as continuous functions of the rotational frequency 
$\omega_\mathrm{rot}$. Detailed expressions of these formulations were given in Ref.~\cite{MM1}. 

Diagonalizations of the PVC Hamiltonian are performed in the model space spanned by 
0 -- 4$\gamma$ basis states as in Ref.~\cite{MM2}. Here $n\gamma$ basis states designate 
$\mathrm{1qp}\otimes(\gamma\text{ or }\bar\gamma)^n$ states. Because there are two types of 
$\gamma$-vibrational phonons, $\gamma$ and $\bar\gamma$, $\mathrm{(1qp)}_{r=\pm i}\otimes\gamma$ 
and $\mathrm{(1qp)}_{r=\mp i}\otimes\bar\gamma$ are possible for the 1$\gamma$ states in 
the $r=\pm i$ sector. Similarly, there are 3 -- 5 types of 2 -- 4$\gamma$ basis states, 
respectively, for each signature. 
The concrete form of eigenstates with $r=-i$ was given by Eq.~(7) in Ref.~\cite{MM2}. That with 
$r=+i$, obtained by interchanging $\mu$ and $\bar\mu$, is given here. 
\begin{equation}
\begin{split}
\left.|\bar{\chi_i}\right\rangle 
&=\sum_{\bar\mu}\psi_i^{(1)}(\bar\mu)\left.a_{\bar\mu}^\dagger|\phi\right\rangle \\
& +\sum_{\bar\mu}\psi_i^{(3)}(\bar\mu\gamma)\left.a_{\bar\mu}^\dagger X_\gamma^\dagger|\phi\right\rangle
  +\sum_{\mu}\psi_i^{(3)}(\mu\bar\gamma)
\left.a_{\mu}^\dagger X_{\bar\gamma}^\dagger|\phi\right\rangle \\
& +\sum_{\bar\mu}\psi_i^{(5)}(\bar\mu\gamma\gamma)\frac{1}{\sqrt{2}}
\left.a_{\bar\mu}^\dagger (X_\gamma^\dagger)^2|\phi\right\rangle
  +\sum_{\bar\mu}\psi_i^{(5)}(\bar\mu\bar\gamma\bar\gamma)\frac{1}{\sqrt{2}}
\left.a_{\bar\mu}^\dagger (X_{\bar\gamma}^\dagger)^2|\phi\right\rangle \\
& +\sum_{\mu}\psi_i^{(5)}(\mu\gamma\bar\gamma)
\left.a_{\mu}^\dagger X_\gamma^\dagger X_{\bar\gamma}^\dagger|\phi\right\rangle \\
& +\sum_{\bar\mu}\psi_i^{(7)}(\bar\mu\gamma\gamma\gamma)\frac{1}{\sqrt{3!}}
\left.a_{\bar\mu}^\dagger (X_\gamma^\dagger)^3|\phi\right\rangle
  +\sum_{\mu}\psi_i^{(7)}(\mu\bar\gamma\bar\gamma\bar\gamma)\frac{1}{\sqrt{3!}}
\left.a_{\mu}^\dagger (X_{\bar\gamma}^\dagger)^3|\phi\right\rangle \\
& +\sum_{\mu}\psi_i^{(7)}(\mu\gamma\gamma\bar\gamma)\frac{1}{\sqrt{2}}
\left.a_{\mu}^\dagger (X_\gamma^\dagger)^2 X_{\bar\gamma}^\dagger|\phi\right\rangle
  +\sum_{\bar\mu}\psi_i^{(7)}(\bar\mu\gamma\bar\gamma\bar\gamma)\frac{1}{\sqrt{2}}
\left.a_{\bar\mu}^\dagger X_{\bar\gamma}^\dagger (X_{\bar\gamma}^\dagger)^2|\phi\right\rangle \\
& +\sum_{\bar\mu}\psi_i^{(9)}(\bar\mu\gamma\gamma\gamma\gamma)\frac{1}{\sqrt{4!}}
\left.a_{\bar\mu}^\dagger (X_\gamma^\dagger)^4|\phi\right\rangle
  +\sum_{\bar\mu}\psi_i^{(9)}(\bar\mu\bar\gamma\bar\gamma\bar\gamma\bar\gamma)\frac{1}{\sqrt{4!}}
\left.a_{\bar\mu}^\dagger (X_{\bar\gamma}^\dagger)^4|\phi\right\rangle \\
& +\sum_{\mu}\psi_i^{(9)}(\mu\gamma\gamma\gamma\bar\gamma)\frac{1}{\sqrt{3!}}
\left.a_{\mu}^\dagger (X_\gamma^\dagger)^3 X_{\bar\gamma}^\dagger|\phi\right\rangle
  +\sum_{\mu}\psi_i^{(9)}(\mu\gamma\bar\gamma\bar\gamma\bar\gamma)\frac{1}{\sqrt{3!}}
\left.a_{\mu}^\dagger X_\gamma^\dagger (X_{\bar\gamma}^\dagger)^3|\phi\right\rangle \\
& +\sum_{\bar\mu}\psi_i^{(9)}(\bar\mu\gamma\gamma\bar\gamma\bar\gamma)\frac{1}{2}
\left.a_{\bar\mu}^\dagger (X_\gamma^\dagger)^2 (X_{\bar\gamma}^\dagger)^2|\phi\right\rangle .
\label{wf}
\end{split}
\end{equation}
In the present work, $\mu$ and $\bar\mu$ run 21 states in the neutron $N_\mathrm{osc}=5$ shell. 

The bands in $^{105}$Mo studied in the present work are the yrast $\nu[532]\,5/2^-$ band and 
corresponding single- and double-$\gamma$-vibrational bands. In the following, qp states 
at finite $\gamma$ and $\omega_\mathrm{rot}$ are also designated by the dominant 
asymptotic states. The lowest qp states with the dominant $\nu[532]\,5/2^-$ component 
are denoted by $\bar\mu=\bar1$ and $\mu=1$ for $r=+i$ and $r=-i$, respectively. 
These are the dominant components of the lowest (zero-phonon) PVC eigenstate in each signature. 
Then the two dominant components of 
the one-phonon PVC eigenstates are $\bar1\otimes\gamma$ and $1\otimes\bar\gamma$ for $r=+i$ 
while $1\otimes\gamma$ and $\bar1\otimes\bar\gamma$ for $r=-i$. Straightforwardly, those 
of the two-phonon in $r=+i$ are $\bar1\otimes\gamma\gamma$, $\bar1\otimes\bar\gamma\bar\gamma$, 
and $1\otimes\gamma\bar\gamma$, while those in $r=-i$ are $1\otimes\gamma\gamma$, 
$1\otimes\bar\gamma\bar\gamma$, and $\bar1\otimes\gamma\bar\gamma$. The sum of the fractions 
of these main components defines the collectivity of calculated eigenstates. 

The parameters entering into the calculation are chosen in a manner similar to the case of 
Refs.~\cite{MM1} and \cite{MM2}. Concretely, the pairing gaps, $\Delta_n=1.05$ MeV and 
$\Delta_p=0.85$ MeV, 
are those widely used for both even- and odd-$A$ nuclides in this mass region~\cite{Mo106,Nb103}. 
The quadrupole deformation, $\epsilon_2=0.3254$, is the same as that adopted for the isobar 
$^{105}$Nb in Ref.~\cite{MM2}. The triaxial deformation $\gamma$ is chosen so that the calculated 
signature splitting $\Delta e'$ between the lowest PVC eigenstates reproduces overall 
features of the observed one. The resulting value, $\gamma=-10^\circ$, is again the same as 
that for $^{105}$Nb. For the quadrupole interaction strengths, examined in the first attempt were 
those which reproduce the observed $\gamma$-vibrational energy $\omega_\gamma$, 0.8121 MeV of 
$^{104}$Mo~\cite{Mo104} or 0.7104 MeV of $^{106}$Mo~\cite{Mo106} or their average, in the RPA 
calculation for the even-even core state at each $\gamma$ as in previous 
calculations~\cite{MSM,Ge,MM1,MM2}. But these were unsuccessful because there exists the 
$\nu[541]\,3/2^-$ 
qp-dominant state slightly higher than the one-phonon states in the PVC calculation and 
the former pushes down the latter. This contradicts the observed situation that the 
$\gamma$-vibrational energy of $^{105}$Mo is higher than those of $^{104}$Mo and $^{106}$Mo, 
see Fig.~10 of Ref.~\cite{Mo105}. 
To reproduce this phenomenologically, adopted are the interaction strengths that give 
$\omega_\gamma=1.0$ MeV in the RPA calculation for the even-even core and consequently 
bring the one-phonon states higher than the $\nu[541]\,3/2^-$-dominant state in the PVC 
calculation at $\omega_\mathrm{rot}=0$. After rotation sets in, calculated 
$\omega_\gamma$ and $\omega_{\bar\gamma}$ in the RPA calculation almost degenerate 
up to $\omega_\mathrm{rot}=0.3$ MeV, then at this rotational frequency, 
the $\gamma$ with $r=+1$ is crossed by the steeply down-slope 
$(\nu[532]\,5/2^-)^2$ state. This crossing produces irregularities in the PVC calculation. 
The configuration with this aligned is the $s$-band state in this mass region and would 
become the yrast in the even-even core when its Routhian reaches 0 at higher 
$\omega_\mathrm{rot}$ although this crossing is blocked in the odd-$A$ system with one of 
the $\nu[532]\,5/2^-$ signature-partner pair is already occupied. Therefore 
irregularities at $\omega_\mathrm{rot}=0.3$ MeV are artifacts of the model and 
should be ignored. After this crossing, $\omega_\gamma$ is slightly larger than 
$\omega_{\bar\gamma}$. 

\section{Results and discussion}

\begin{figure}[htbp]
\includegraphics[width=6cm]{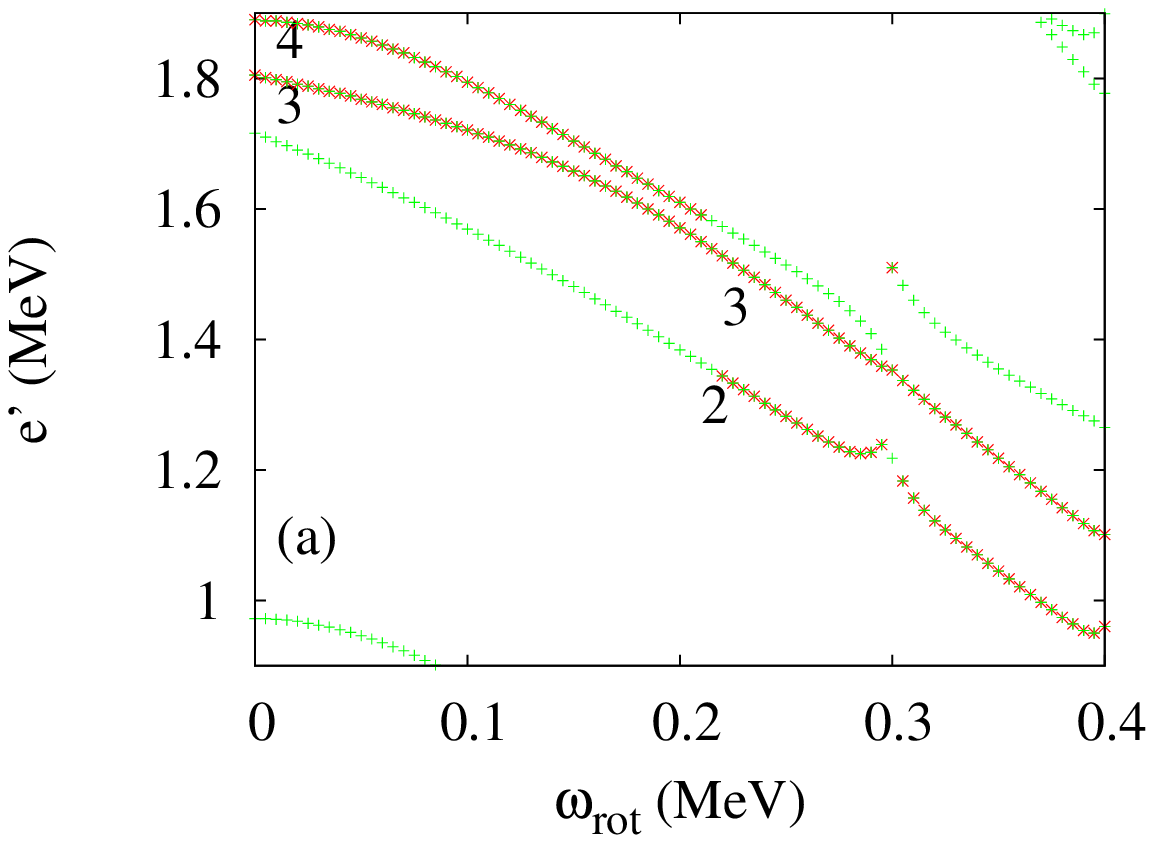}
\includegraphics[width=6cm]{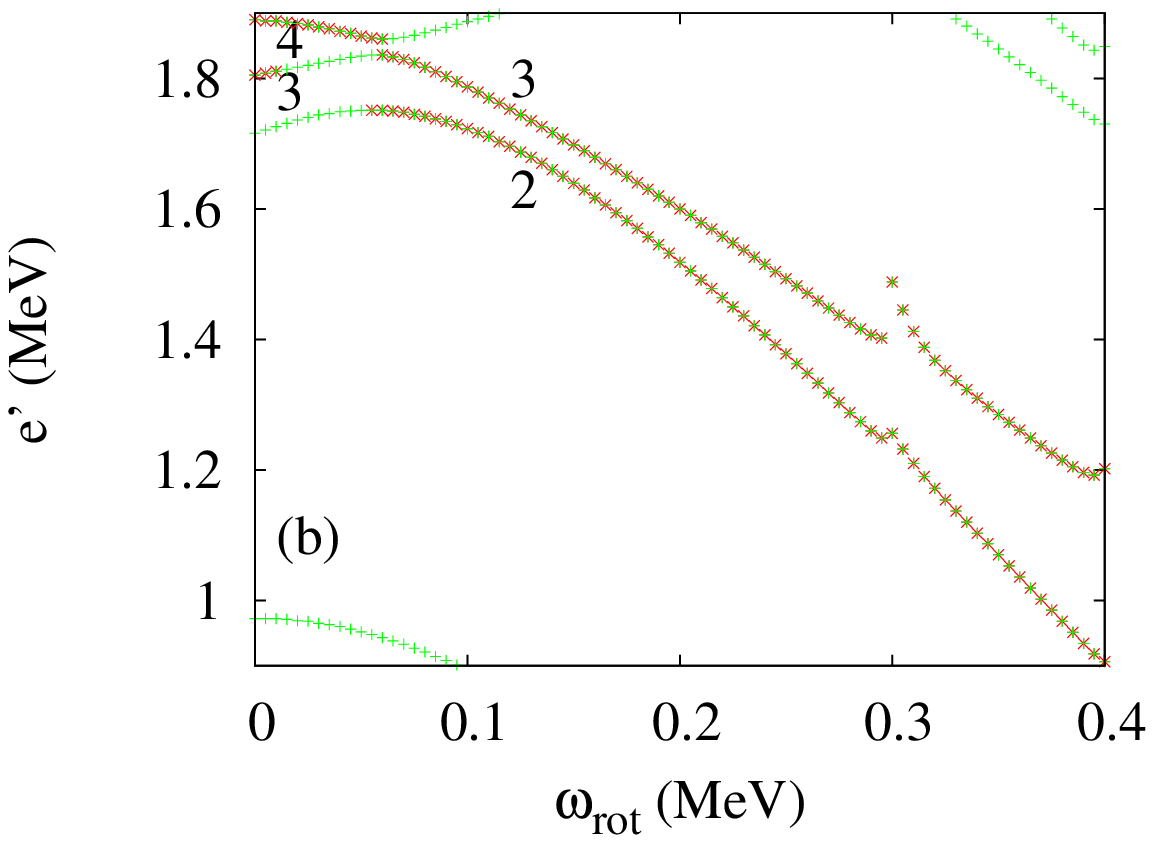}
 \caption{(Color online) Routhians of all calculated PVC states in the region 
of one-phonon bands are shown by green $+$s,(a) in the $r=+i$ sector and (b) in the $r=-i$ 
sector. 
Those with more than 50\% collective fraction are emphasized by red $\times$s. 
The labels attached designate the numbers, $i$ in Eq.~(\ref{wf}) in the present article for $r=+i$ 
and in Eq.~(7) in Ref.~\cite{MM2} for $-i$, enumerated from the lowest. Note that the lowest 
state seen only in small $\omega_\mathrm{rot}$ in each figure is the yrast zero-phonon state.}
 \label{fig1}
\end{figure}

The Routhians in the one-phonon region are shown in Figs.~\ref{fig1} (a) for $r=+i$ and 
(b) for $r=-i$. There are three eigenstates around $e'=1.8$ MeV at $\omega_\mathrm{rot}\sim0$ 
in both signature sectors. 
If the static triaxial deformation that mixes the $K$ quantum number is ignored, the second, 
third, and fourth states are the $\nu[541]\,3/2^-$, the $K=\Omega-2=1/2$ $\gamma$ vibration, 
and the $K=\Omega+2=9/2$ $\gamma$ vibration, respectively, hereafter $j$ and $\Omega$ are 
the single-particle angular momentum and its projection to the $z$ axis. 

In the single-$j$ approximation effective for high-$j$ cases, the favored state ($r=+i$ 
in the present case) is written as 
$|\mathrm{f}_\Omega\rangle=\frac{1}{\sqrt{2}}\left(|\Omega\rangle+|-\Omega\rangle\right)$ 
while the unfavored state ($r=-i$ in the present case) is 
$|\mathrm{u}_\Omega\rangle=\frac{1}{\sqrt{2}}\left(|\Omega\rangle-|-\Omega\rangle\right)$ 
at $\omega_\mathrm{rot}=0$ aside from overall phases. Therefore those with $\Omega=1/2$ 
have diagonal matrix elements of $J_x$ and consequently their Routhians, eigenvalues of 
$h'=h-\omega_\mathrm{rot}J_x$, split as soon as rotation sets in. 
This splitting propagates to other states through the $\Omega$ mixing at finite 
$\omega_\mathrm{rot}$. This is the basic mechanism of the signature splitting. 
In addition, when triaxial deformation, represented 
by the term proportional to $Q_2^{(+)}=\frac{1}{\sqrt{2}}\left(Q_{22}+Q_{2-2}\right)$ in 
$h$, and/or couplings to the $\gamma$ vibration exist, $\Omega=3/2$ states also split through 
$\langle3/2|Q_{22}|-1/2\rangle\times\langle-1/2|J_x|1/2\rangle\times\langle1/2|Q_{22}|-3/2\rangle$, 
for example. This is the reason why both the noncollective second and the collective third states 
exhibit finite alignments $=-\mathrm{d}e'/\mathrm{d}\omega_\mathrm{rot}$ at 
$\omega_\mathrm{rot}=0$ and interact with each other at finite $\omega_\mathrm{rot}$. 
Because the mutual repulsion extends over wide range of $\omega_\mathrm{rot}$, 
the third (originally lower one-phonon) state is pushed up and interacts 
also with the fourth (originally upper one-phonon) state. Then at $\omega_\mathrm{rot}>0.2$ MeV
the third state carries the character of the upper one-phonon state in $r=+i$ whereas the effect 
of the noncollective state is only local in $r=-i$. 

Aside from this perturbation from the noncollective state, the collective states show 
the pattern, common to the previous cases of $^{103}$Nb and $^{105}$Nb, 
that the lower one-phonon state has $K=\Omega-2$ with a finite alignment 
(positive for the favored and negative for the unfavored) and the upper one has $K=\Omega+2$ 
with negligible alignment for small $\omega_\mathrm{rot}$, see Fig.~3(a) in Ref.~\cite{MM2}. 
For example, in the $r=-i$ sector, both of them are dominantly composed of the 
$1\otimes\gamma$ and the $\bar1\otimes\bar\gamma$ basis states. Two orthogonal combinations 
of them with similar magnitudes form the eigenstates with $K=\Omega-2$ and $\Omega+2$ as 
discussed in Ref.~\cite{MM2}. This pattern is preserved in the whole calculated range of 
$\omega_\mathrm{rot}$ in the cases of $^{103}$Nb and $^{105}$Nb with the $\pi g_{9/2}$ odd 
particle with small signature splittings studied there. 

In the present case of $^{105}$Mo with the higher-$j$, $\nu h_{11/2}$, odd particle, 
in contrast, stronger Coriolis interactions change their characters as $\omega_\mathrm{rot}$ 
increases. Actually slopes of their Routhians are as twice as those of 
$^{103}$Nb and $^{105}$Nb. Because the signature splitting between the 1 and the $\bar1$ basis 
states is significantly larger than that between $\gamma$ and $\bar\gamma$, 
the $\bar1\otimes\bar\gamma$ component becomes dominant in the lower one-phonon state while 
the $1\otimes\gamma$ component becomes dominant in the upper one in the $r=-i$ sector, 
for example. 

\begin{figure}[htbp]
\includegraphics[width=6cm]{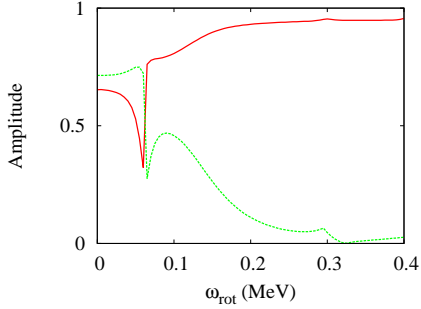}
\caption{(Color online) Amplitudes of the dominant components 
$|\psi_i^{(3)}(1\gamma)|$ (red solid) and $|\psi_i^{(3)}(\bar 1 \bar\gamma)|$ (green dashed) 
in the upper ($i=3$ at $\omega_\mathrm{rot}>0.06\mathrm{MeV}$) one-phonon band in the 
$r=-i$ sector.}
 \label{fig2}
\end{figure}

\begin{figure}[htbp]
\includegraphics[width=10cm]{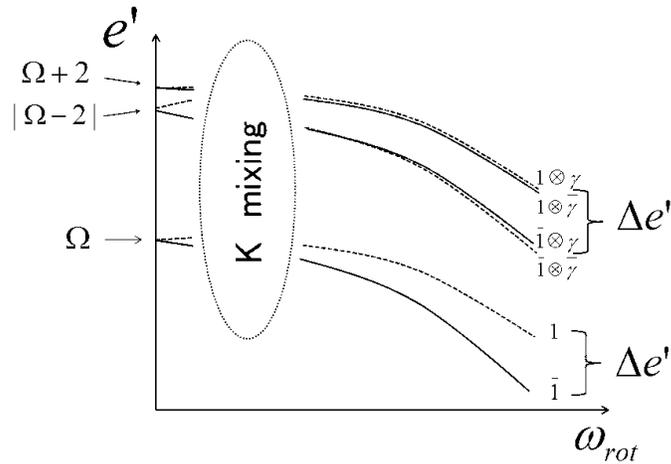}
 \caption{Schematic drawing of the character change of the zero- and one-phonon bands from 
the $K$ scheme to the signature scheme caused by the Coriolis $K$ mixing.}
 \label{fig3}
\end{figure}

This character change caused by the Coriolis $K$ mixing can be 
clearly seen in the wave function shown in Fig.~\ref{fig2} for $r=-i$. 
Aside from the abrupt change brought about by the crossing with the noncollective 
state at around $\omega_\mathrm{rot}=0.06$ MeV, the $1\otimes\gamma$ component 
becomes dominant in the upper one-phonon state at high $\omega_\mathrm{rot}$. 
The schematic behavior of the collective states is depicted in Fig.~\ref{fig3}. 
This figure clearly explains the reason why the observed one-phonon band does not exhibit 
signature splitting in contrast to the yrast zero-phonon band: The observed one-phonon 
band with the $K=\Omega+2$ band head does not show signature splitting at 
$\omega_\mathrm{rot}\sim0$ because of the high $K$ as argued in Ref.~\cite{MM2}. 
And at finite $\omega_\mathrm{rot}$, the $r=+i$ and $r=-i$ sequences of the band consist 
dominantly of the $1\otimes\bar\gamma$ and $1\otimes\gamma$, respectively, 
hence the splitting between them is essentially the difference 
$\omega_\gamma-\omega_{\bar\gamma}$, 
which is much smaller than the splitting in the yrast zero-phonon band. 
When the $r=-i$ is favored as in the $g_{9/2}$ and $i_{13/2}$ cases, 
1 and $\bar1$ in the figure should be interchanged. 

\begin{figure}[htbp]
\includegraphics[width=6cm]{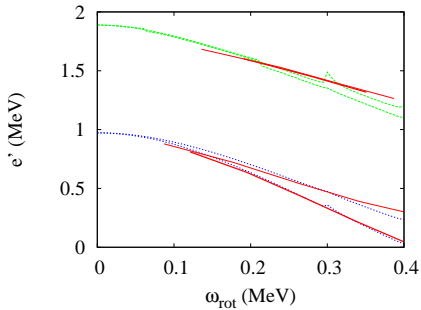}
 \caption{(Color online) Routhians of calculated zero-phonon (blue dotted) and upper 
($K=\Omega+2$) one-phonon (green dashed) bands are compared with the corresponding data 
(red solid) converted to the rotating frame by using the Harris parameters 
$\mathcal{J}_0=18.08$ MeV$^{-1}$ and $\mathcal{J}_1=43.21$ MeV$^{-3}$ 
that fit the yrast band of $^{106}$Mo~\cite{Mo104}. 
Thick and thin curves designate the $r=+i$ and $-i$, respectively.}
 \label{fig4}
\end{figure}

The calculated zero- and one-phonon states are compared quantitatively with the observed ones. 
The reference rotating frame to which the data are converted is determined in two steps. 
First the Harris parameters are determined to fit the ground band up to $10^+$ of 
$^{106}$Mo as $\mathcal{J}_0=18.08$ MeV$^{-1}$ and $\mathcal{J}_1=43.21$ MeV$^{-3}$. 
In the second step the origin of this frame must be shifted in order to be compared with 
states in the odd-$A$ system. Usually this overall shift is given by the pairing gap, 
in the present case $\Delta_n=1.05$ MeV, in the cranking calculation. In the present 
calculation, however, the cranking model is extended to the PVC model and this produces 
a downward shift of the Routhian of the lowest state, 0.078 MeV at $\omega_\mathrm{rot}=0$. 
Accordingly the overall shift is determined to be 0.972 MeV. 
The result is shown in Fig.~\ref{fig4}. This shows that the zero- and one-phonon states are 
reproduced well. 

\begin{figure}[htbp]
\includegraphics[width=6cm]{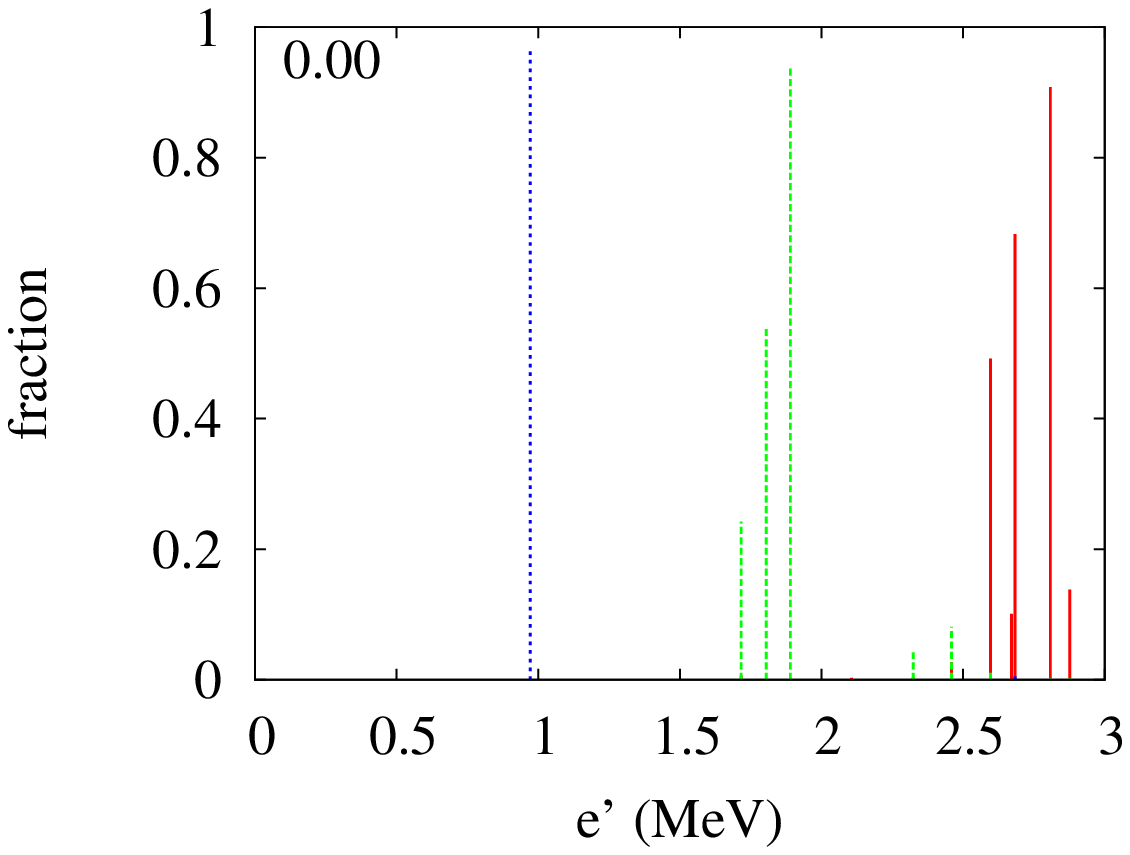}
\includegraphics[width=6cm]{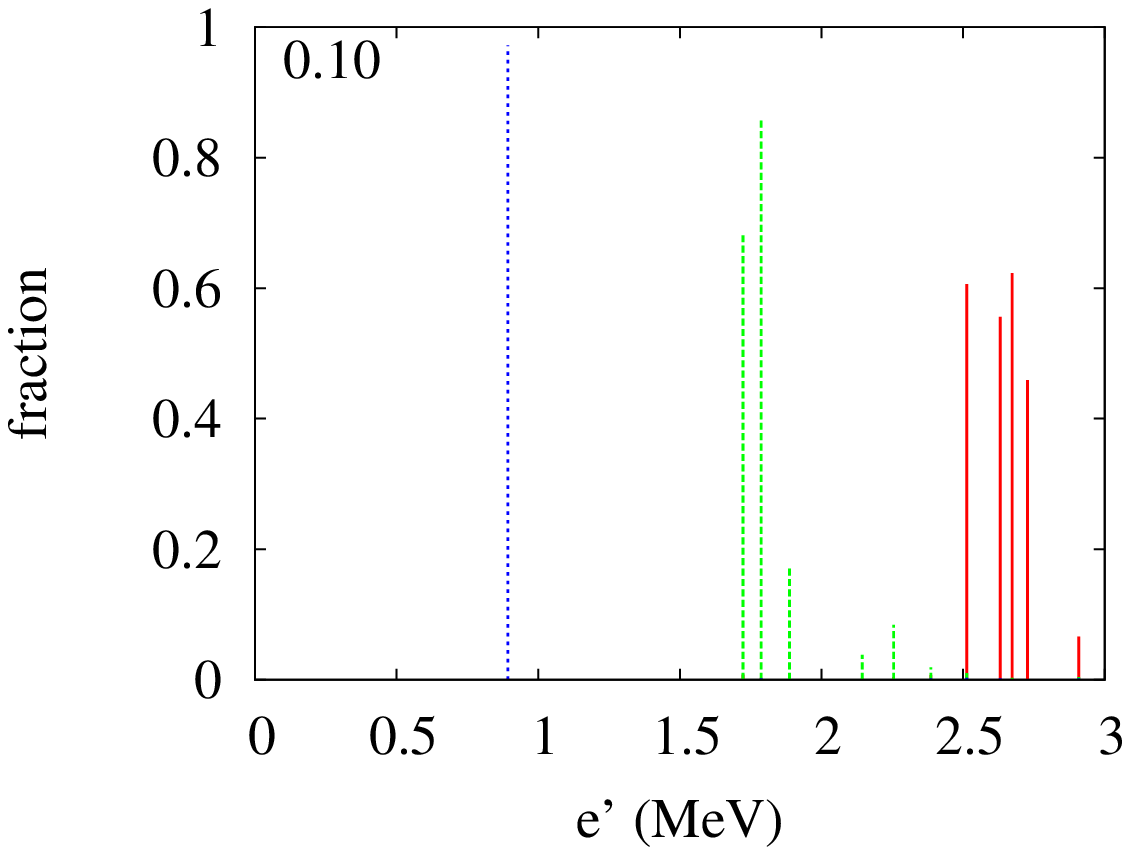}
\includegraphics[width=6cm]{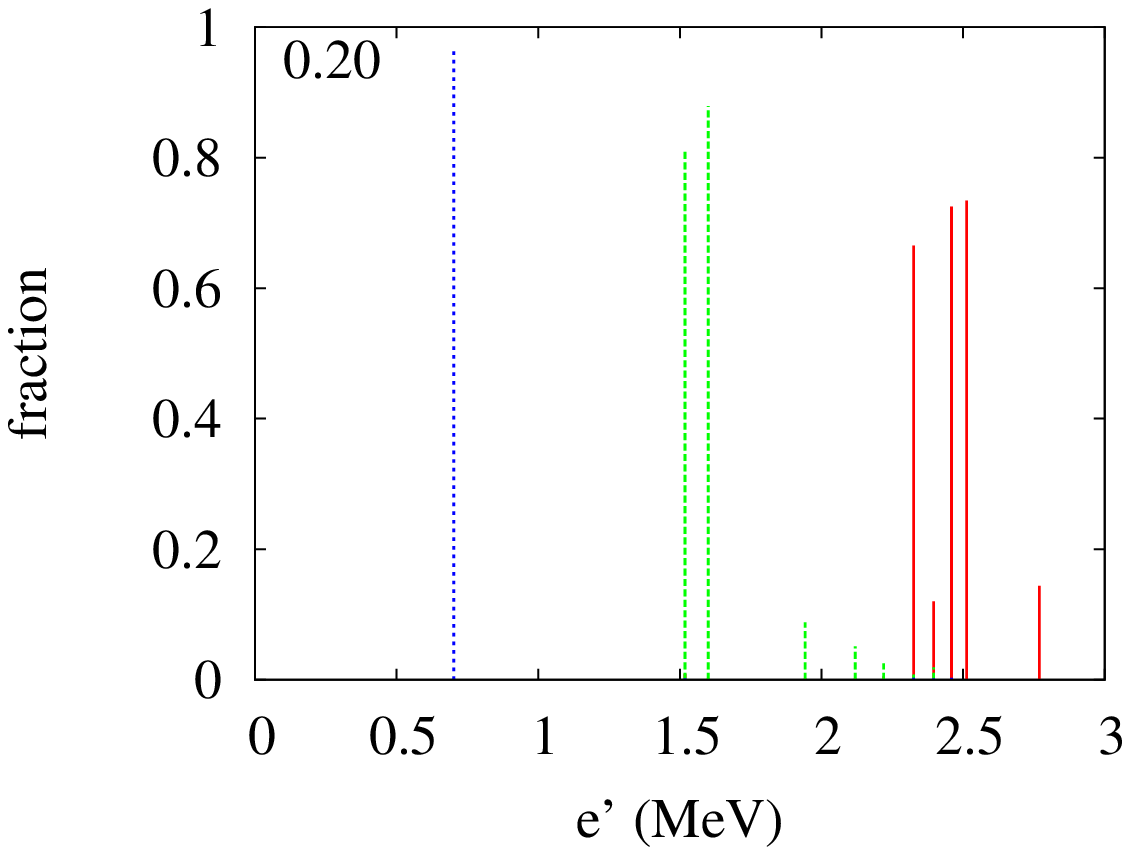}
\includegraphics[width=6cm]{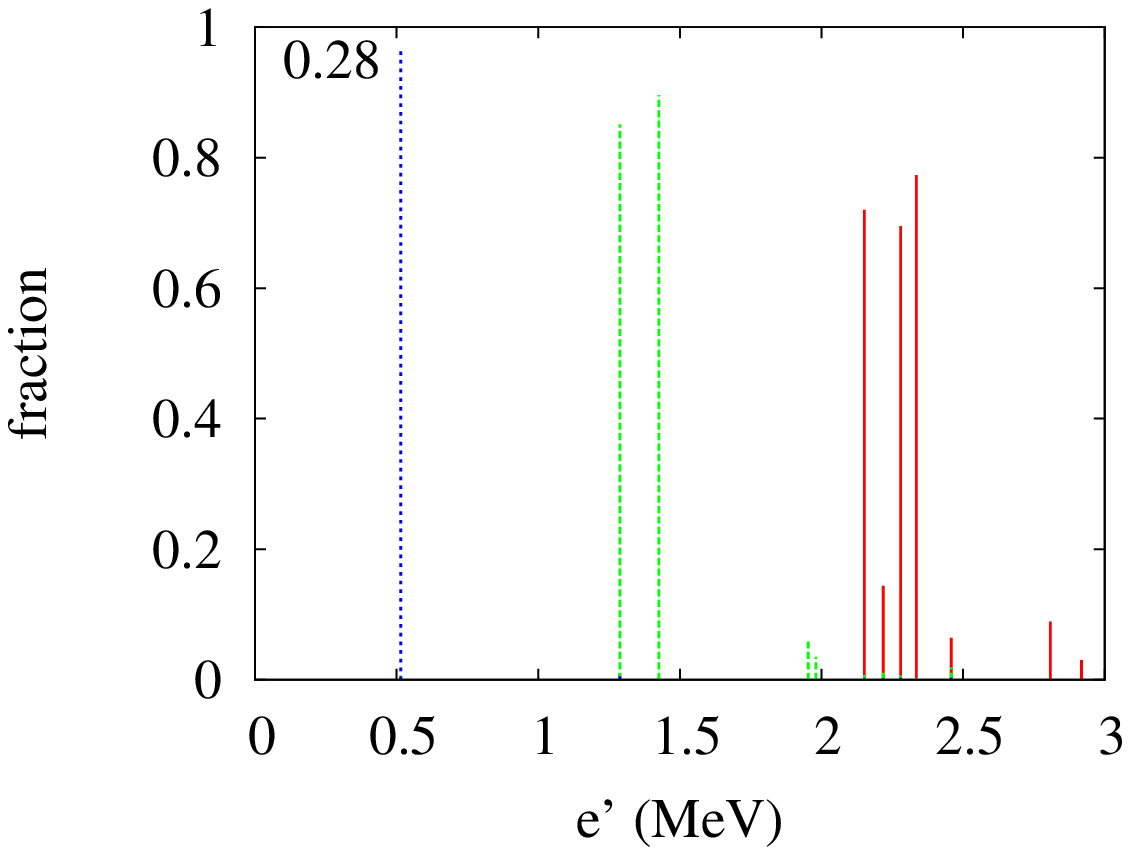}
 \caption{(Color online) Distribution of the collective fraction (probability in the 
wave function) of the zero-, one-, and two-phonon components in the $r=-i$ sector, 
$|\psi^{(1)}(1)|^2$ (blue dotted), 
$|\psi^{(3)}(1\gamma)|^2+|\psi^{(3)}(\bar 1 \bar\gamma)|^2$ (green dashed), 
and $|\psi^{(5)}(1\gamma\gamma)|^2+|\psi^{(5)}(1\bar\gamma\bar\gamma)|^2+
 |\psi^{(5)}(\bar 1 \gamma\bar\gamma)|^2$ (red solid), respectively, 
at $\omega_\mathrm{rot}=0$, 0.1, 0.2, and 0.28 MeV.}
 \label{fig5}
\end{figure}

Now I proceed to the two-phonon states. 
How their calculated collectivity distributes is shown in Fig.~\ref{fig5} for $r=-i$ 
together with the zero- and one-phonon states at four $\omega_\mathrm{rot}$s. 
These figures show that the locations of the one-phonon and two-phonon states are harmonic 
at least their centers of the collective strength are seen. 
Experimentally, however, an anharmonicity, $E_{2\gamma}/E_{1\gamma}=1.76$, 
was reported in contrast to the harmonic spectra in the adjacent even-even isotopes, 
$^{104}$Mo~\cite{Mo104} and $^{106}$Mo~\cite{Mo106}. Note here that an anharmonic $E_{2\gamma}$ 
in $^{106}$Mo was cited in Ref.~\cite{Mo105} referring to Ref.~\cite{Mo106_2} but 
it is a misciting of the band-head energy of another band. Reference~\cite{Mo106_2} reported 
the same harmonic value. These data suggest an unknown mechanism 
to produce anharmonicity proper to odd-$A$ systems, different from the one discussed for 
even-even nuclei from a microscopic theoretical point of view in Ref.~\cite{MatsuoMatsu}. 

\begin{figure}[htbp]
\includegraphics[width=6cm]{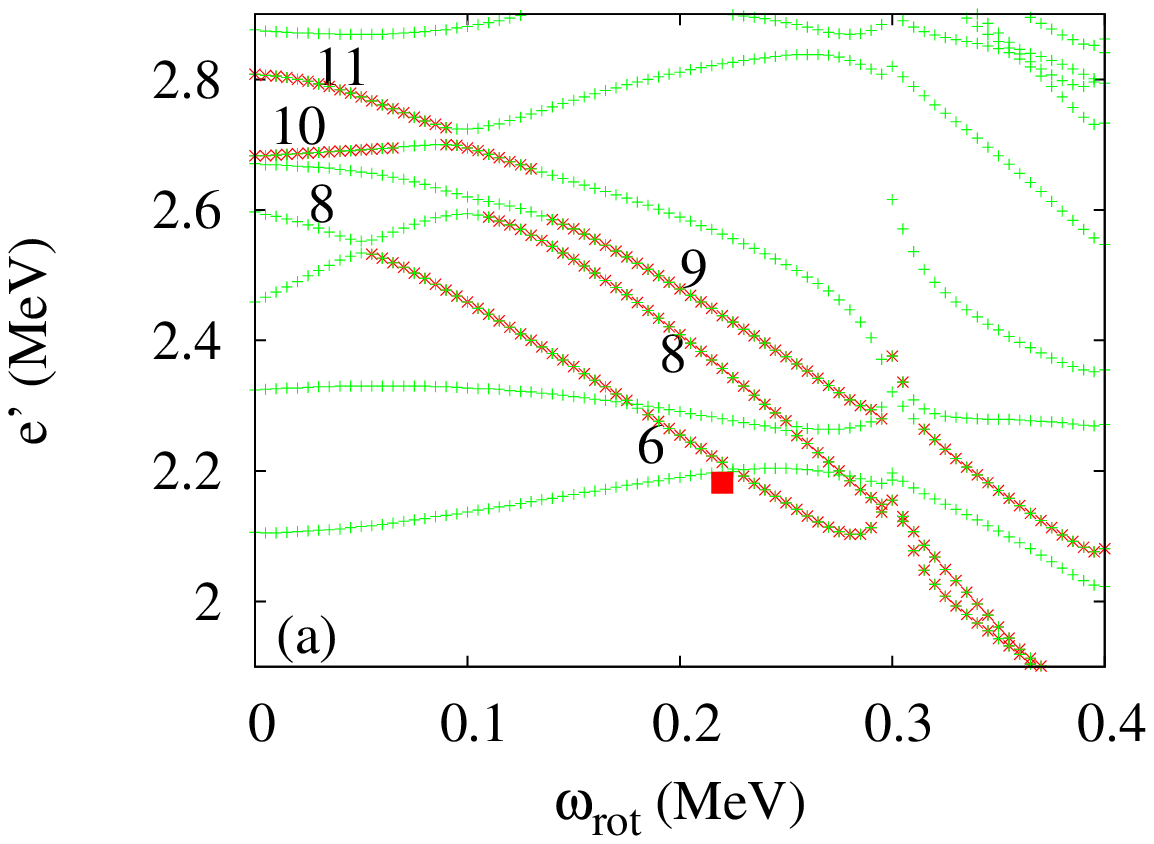}
\includegraphics[width=6cm]{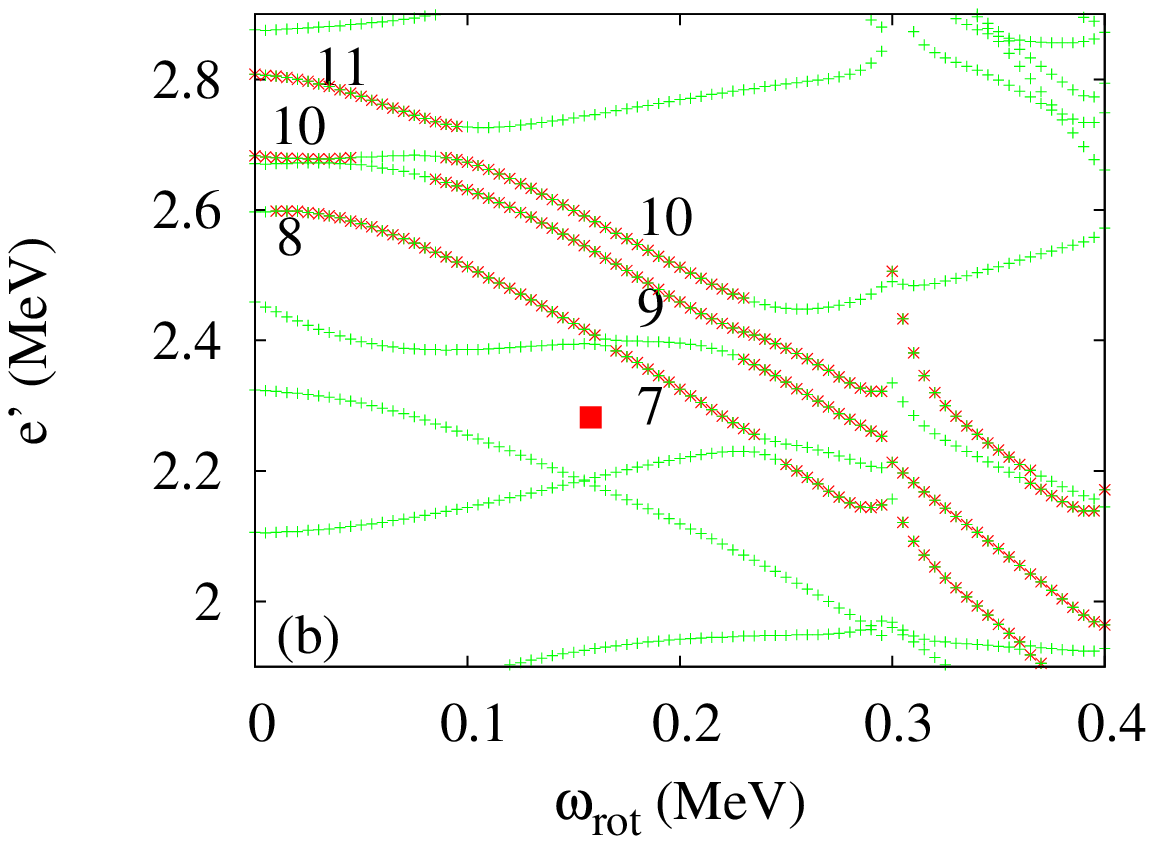}
 \caption{(Color online) The same as Fig.~\ref{fig1} but in the region of two-phonon bands.
Corresponding data (red filled square) converted to the rotating frame are also shown.}
 \label{fig6}
\end{figure}

More closely, their Routhians are shown in Figs.~\ref{fig6} (a) for $r=+i$ and (b) for $r=-i$ 
as continuous functions of $\omega_\mathrm{rot}$. As in the cases of $^{103}$Nb and $^{105}$Nb, 
three sequences keep collectivity in the whole calculated range of $\omega_\mathrm{rot}$ 
although they are located in the region with larger level density. 
The lower one is of $K=|\Omega-4|=3/2$, the medium one is of $K=\Omega=5/2$, and 
the upper one is of $K=\Omega+4=13/2$ at $\omega_\mathrm{rot}\sim0$. 
Experimentally, the band head was assigned as $K=13/2$ and only two states in each 
signature were observed. They are represented by a point in the Routhian diagram for 
each signature as indicated by red filled squares in Fig.~\ref{fig6}. 

The locations of the highest peaks of each phonon number in Fig.~\ref{fig5} for 
$\omega_\mathrm{rot}=0$ do not 
account for the observed anharmonicity at the band head. 
The calculated Routhians of the zero-phonon, (upper) one-phonon, and (upper) two-phonon 
states at $\omega_\mathrm{rot}=0$ are (1.050, 2.050, 3.050) MeV without PVC (not shown) while 
(0.972, 1.890, 2.808) MeV with PVC, respectively. 
However, at finite $\omega_\mathrm{rot}$ where actual transitions are observed, 
the Coriolis effect is significant in the present high-$j$ case and calculated states look 
in the signature scheme with mixed $K$. 
Inspection of the wave functions indicates that the dominant components are 
$\bar1\otimes\gamma\gamma$ for the lower, 
$\bar1\otimes\bar\gamma\bar\gamma$ for the medium, and $1\otimes\gamma\bar\gamma$ for the 
upper in the $r=+i$ sector, while $\bar1\otimes\gamma\bar\gamma$ for the lower, 
$1\otimes\bar\gamma\bar\gamma$ for the medium, and $1\otimes\gamma\gamma$ for the 
upper in the $r=-i$ sector. Again this proves that the order of the collective states 
is ruled by the signature splitting between the 1 and $\bar1$ basis states as in the 
one-phonon states in Fig.~\ref{fig3}. This indicates that their connection to the band head 
with fairly pure $K$ is not trivial. 

Then a natural criterion to choose the state to be identified with (the main component of) 
the observed one is the largest collectivity, as argued in the three-phonon-candidate cases 
in Ref.~\cite{MM2}, because it is generally expected that the most collective sequence 
connected by strong $E2$ transitions would be observed. 
Actually the upper one, identified with the observed one at the band head, is the 
most collective at $\omega_\mathrm{rot}=0$ because its high-$K$ property prevents 
it from mixing with other states. But three sequences become to share 
similar collectivities as soon as rotation sets in. 
As in the three-phonon case, the state with the highest $K$ 
effectively has the highest $j$ and accordingly feels the strongest 
Coriolis force as $\omega_\mathrm{rot}$ increases. Then it is expected that band members 
that have large overlaps with the most collective eleventh state with the highest $K$ 
at $\omega_\mathrm{rot}\sim0$ 
would be energetically lowered with reducing $K$ as $\omega_\mathrm{rot}$ increases and appear 
at lower Routhians through bandcrossing(s).
The observed transitions are located near to ($r=+i$) or even lower than ($r=-i$) the lowest 
calculated two-phonon states. 
However, lowering of the location of collectivity to these lowest states in the present 
calculation is insufficient in contrast to the three-phonon case although $j$ is larger. 
Therefore it is difficult to establish the mapping between them. 

\section{Conclusions}

To conclude, the yrast (zero-phonon) $\nu[532]\,5/2^-$, the one- and two-phonon 
$\gamma$ vibrational bands in $^{105}$Mo have been calculated in the particle-vibration 
coupling model based on the cranking model and the random-phase approximation paying 
attention to the rotational effects on the spectra in comparison to the lower-$j$ 
cases of $^{103}$Nb and $^{105}$Nb studied in the previous works. 

The zero- and one-phonon bands have been reproduced well. In particular, the rotational 
character change from the $K$ scheme to the signature scheme through the Coriolis 
$K$ mixing in the one-phonon states is stressed. 
This naturally accounts for the reason why the observed one-phonon band does not exhibit 
signature splitting in contrast to the yrast zero-phonon band. 
A specific feature of the data is that the two-phonon 
states show anharmonicity in the spectra that is absent in $^{103}$Nb, $^{105}$Nb, $^{104}$Mo, 
and $^{106}$Mo. This fact suggests that there exists an unknown mechanism to produce 
anharmonicity proper to high-$j$ odd-$A$ nuclei. 
The particle-vibration coupling pushes down them but the result is still harmonic 
with a slightly reduced interval at $\omega_\mathrm{rot}=0$. 
A possibility applicable to finite $\omega_\mathrm{rot}$, 
inspired from the previous calculation for the three-phonon-candidate 
states in $^{103}$Nb and $^{105}$Nb, is that the continuation of the highest-lying 
two-phonon state with the highest $K$ would be energetically lowered by a strong Coriolis force. 
In the present calculation, however, its lowering is insufficient.


\begin{thebibliography}{99}

\bibitem{BABH}
 B. Bally, B. Avez, M. Bender, and P. -H. Heenen, Phys. Rev. Lett. {\bf 113}, 162501 (2014).

\bibitem{Er168}
 W. F. Davidson et al., J. Phys. {\bf G7}, 455; 843 (1981).

\bibitem{Er166_1}
 C. Fahlander et al., Phys. Lett. {\bf B388}, 475 (1996).

\bibitem{Er166_2}
 P. E. Garrett et al., Phys. Rev. Lett. {\bf 78}, 4545 (1997).

\bibitem{Dy164}
 F. Corminboeuf et al., Phys. Rev. {\bf C56}, R1201 (1997). 

\bibitem{Th232_1}
 W. Korten et al., Phys. Lett. {\bf B317}, 19 (1993).

\bibitem{Mo106}
 A. Guessous et al., Phys. Rev. Lett. {\bf 75}, 2280 (1995).

\bibitem{Mo104}
 A. Guessous et al., Phys. Rev. {\bf C53}, 1191 (1996).

\bibitem{Nd138}
 H. J. Li et al., Phys. Rev. {\bf C87}, 057303 (2013).

\bibitem{MM1}
 M. Matsuzaki, Phys. Rev. {\bf C83}, 054320 (2011).

\bibitem{Mo105}
 H. B. Ding et al., Phys. Rev. {\bf C74}, 054301 (2006).

\bibitem{Nb103}
 J. -G. Wang et al., Phys. Lett. {\bf B675}, 420 (2009).

\bibitem{Tc107}
 L. Gu et al., Chin. Phys. Lett. {\bf 26}, 092502 (2009).

\bibitem{SBSP}
 J. A. Sheikh, G. H. Bhat, Y. Sun, and R. Palit, Phys. Lett. 
 {\bf B688}, 305 (2010).

\bibitem{Nb105}
 H. J. Li et al., Phys. Rev. {\bf C88}, 054311 (2013).

\bibitem{MM2}
 M. Matsuzaki, Phys. Rev. {\bf C90}, 044313 (2014).

\bibitem{SN}
 Y. R. Shimizu and T. Nakatsukasa, Nucl. Phys. {\bf A611}, 22 (1996). 

\bibitem{MSM}
 M. Matsuzaki, Y. R. Shimizu and K. Matsuyanagi, Prog. Theor. 
 Phys. {\bf 77, 1302} (1987); {\it ibid.} {\bf 79}, 836 (1988).

\bibitem{Ge}
 G. Gervais et al., Nucl. Phys. {\bf A624}, 257 (1997). 

\bibitem{Mo106_2}
 R. Q. Xu et al., Chin. Phys. Lett. {\bf 19}, 180 (2002).

\bibitem{Mo106_3}
 H. Hua et al., Phys. Rev. {\bf C69}, 014317 (2004).

\bibitem{MatsuoMatsu}
 M. Matsuo and K. Matsuyanagi, Prog. Theor. Phys. {\bf 74}, 1227 (1985); 
{\it ibid.} {\bf 76}, 93 (1986); {\it ibid.} {\bf 78}, 591 (1987).

\end{thebibliography}
\end{document}